\documentclass{jfm}
\usepackage{graphicx}
\usepackage{epstopdf, epsfig}
\usepackage{color, soul}
\usepackage{esint}

\shorttitle{Cavity collapse near slot geometries}
\shortauthor{E. D. Andrews, D. Fern\'andez Rivas and I. R. Peters}

\title{Cavity collapse near slot geometries}

\author{Elijah D. Andrews\corresp{\email{e.d.andrews@soton.ac.uk}}\aff{1},
 David Fern\'andez Rivas\aff{2}
 \and Ivo R. Peters\aff{1}}

\affiliation{\aff{1}Faculty of Engineering and Physical Sciences, University of Southampton, Southampton SO17 1BJ, UK
\aff{2}Mesoscale Chemical Systems Group, MESA+ Institute, TechMed Centre and Faculty of Science and Technology, University of Twente,
P.O. Box 217, 7500AE Enschede, The Netherlands}

\begin{document}
\maketitle

\begin{abstract}
The collapse of a gas or vapour bubble near a solid boundary produces a jet directed towards the boundary. High surface pressure and shear stress induced by this jet can damage, or clean, the surface. More complex geometries will result in changes in collapse behaviour, in particular the direction of the jet. The majority of prior research has focused on simple flat boundaries or limited cases with analytic solutions. We numerically and experimentally investigate how a slot in a flat boundary affects the jet direction for a single bubble. We use a boundary element model to predict how the jet direction depends on key geometric parameters and show that the results collapse to a single curve when the parameters are normalised appropriately. We then experimentally validate the predictions using laser-induced cavitation and compare the experimental results to the predicted dependencies. This research reveals a tendency for the jet to be directed away from a slot and shows that the jet direction is independent of slot height for slots of sufficient height.
\end{abstract}

\section{Introduction}
The collapse of a gas or vapour bubble near a solid boundary forms a liquid jet that impinges on the boundary \citep{Plesset1971}. The impinging jet can damage, or clean, the boundary surface due to the pressure and wall shear stress induced at the boundary \citep{Dijkink2008a, Koukouvinis2018}. In addition to the jet impingement, shock waves emitted during collapse events may contribute to the effects on the boundary, however it is not clear in which regimes the jet or shock waves are more dominant \citep{VanWijngaarden2016}. Studying the collapse of single bubbles is a valuable tool for understanding the mechanics involved as the effects of the jet are the result of individual bubble collapses \citep{Benjamin1966}. An understanding of individual bubble collapses can also inform investigation of cavitation erosion caused by many individual collapse events building up over time \citep{FernandezRivas2013}. Some common areas in which these bubbles occur are cavitation damage \citep{Sreedhar2017}, hydraulic systems \citep{Luo2016}, and ultrasonic cleaning \citep{Verhaagen2016}. 

Developing a better understanding of the effects of cavitation collapse can increase the lifespan of components such as ship propellers \citep{VanTerwisga2007} and tidal turbines \citep{Kumar2010} or determine the effectiveness of ultrasonic cleaning for complex geometries \citep{Verhaagen2016a, Reuter2017}. There are also numerous applications in biomedical fields such as reducing tissue damage during surgery \citep{Palanker2002}, investigating mechanisms of cell death in cases where cavitation could be used for drug delivery \citep{Dijkink2008}, using cavitation to facilitate needle-free injections \citep{OyarteGalvez2020}, and studying the contribution of cavitation in traumatic brain injuries \citep{Canchi2017} where cavitation is induced by high accelerations \citep{Pan2017}. More novel applications of cavitation include producing high speed liquid jets at small scales \citep{Karri2012}, using cavitation as an ice-breaking mechanism \citep{Cui2018}, and understanding biological mechanisms such as those employed by the snapping shrimp \citep{Versluis2000, Shimu2019}.

The majority of prior experimental research has focused on the jet dynamics near simple flat boundaries \citep{Benjamin1966, Plesset1971, Kucera1990, Dijkink2008a, Supponen2016} or cases with limited complexity such as axisymmetric boundaries \citep{TOMITA2002} or parallel boundaries \citep{Han2018, Gonzalez-Avila2020}. Although understanding how bubbles behave near simple geometries is important, in reality geometries have many more complex features such as corners, indentations, slots, and surface imperfections. For this reason further investigation of complex geometries is important.

Some more complex geometries have been studied, such as parallel boundaries closed at one end \citep{Brujan2019}, semi-infinite boundaries \citep{Kucera1990}, near combinations of a free surface and an inclined flat boundary \citep{Zhang2017}, inside a set of concave corners \citep{Kucera1990, Brujan2018, PhysRevFluids.3.081601}, and inside rectangular and triangular channels \citep{Molefe2019}. There have also been investigations into related phenomena such as the behaviour of ultrasonically driven bubble clouds near larger bubbles trapped on pit geometries \citep{Stricker2013} and bubble collapse dynamics in microfluidic systems of channels with various shapes \citep{Zwaan2007}.

A liquid jet impinging on a boundary can be characterised by its strength and direction. In this paper we investigate the effect of a slot in a flat plate (shown in figure \ref{fig:slot_schematic}a) on the jet direction. Slot geometries are common and some other geometries in flat surfaces could be modelled as a series of slots. Examples of geometries that could be approximated by slots are scratches in flat surfaces, trenches in semiconductor manufacturing, and various 3D printed objects. These slots could impact how well the surfaces are cleaned or cause concentrations of cavitation damage.

The paper is structured as follows. The problem is defined in section \ref{sec:problem definition} and some qualitative predictions are made. The experimental procedure is outlined in section \ref{sec:Experimental Method} and the numerical method is defined in section \ref{sec:Numerical Method}, including some key mathematical derivations. The numerical method is then employed in section \ref{sec:Numerical Results} to make some quantitative predictions of the jet direction. Experimental results are presented in section \ref{sec:Experimental Results}, and subsequently compared to the numerical predictions in section \ref{sec:Comparison}.

\section{Problem Definition}
\label{sec:problem definition}
\begin{figure}
\centerline{\includegraphics{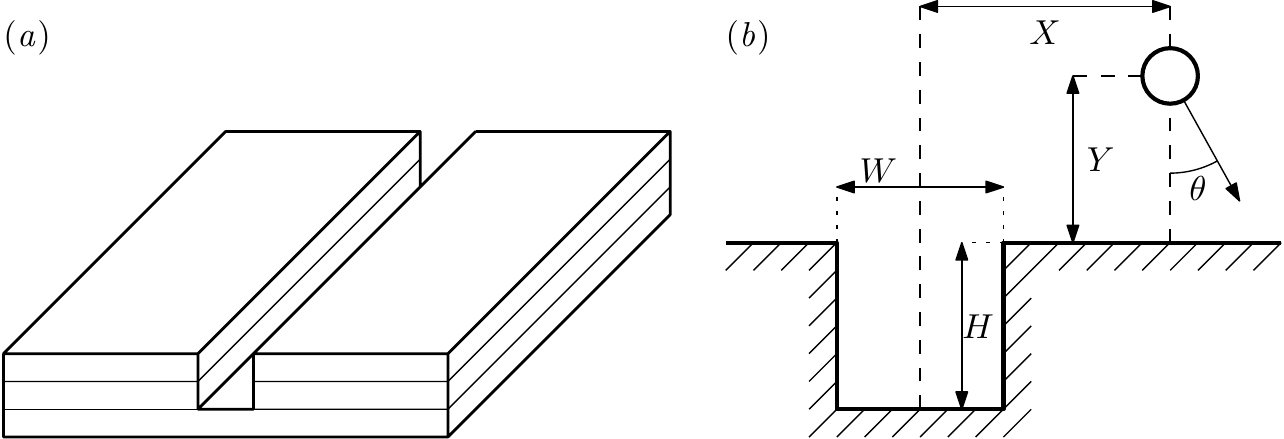}}
\caption{(\textit{a}) Schematic view of a slot in the surface of a flat boundary constructed from layered acrylic. (\textit{b}) The parameters defining a slot, bubble position, and jet direction.}
\label{fig:slot_schematic}
\end{figure}

We define a slot as a rectangular channel in a surface, as shown in figure \ref{fig:slot_schematic}a. The slot has width $W$ and height $H$. A bubble is positioned horizontally at a distance $X$ from the slot centre and vertically at a distance $Y$ from the boundary surface. The bubble radius can be neglected when considering only the jet direction, which is discussed further in section \ref{sec:average surface velocity}. The jet direction is measured anticlockwise from the downwards direction such that a positive angle is towards the right side of the slot. These definitions are shown in figure \ref{fig:slot_schematic}b. The jet angle is a function of the other four parameters, as defined in equation \ref{eq:theta_functions}. From dimensional analysis, this function can be reduced to a function of the three non-dimensional variables $x = 2X/W$, $y = Y/W$, and $h = H/W$. The horizontal position of the bubble is normalised with respect to half of the width of the slot so that the bubble is directly above the edge of the slot at $x = 1$. The vertical position of the bubble is normalised with respect to the width as this provides the most versatility. A slot with an infinite height can still be regarded as a slot, whereas a slot with infinite width is no longer a slot. Normalizing with only the width allows the non-dimensional vertical position to retain relevance for all geometries that could be considered as slots. The jet angle, defined in both forms, is thus
\begin{equation}
\theta = f(X, Y, H, W) = g(x, y, h).
\label{eq:theta_functions}
\end{equation}

\begin{figure}
\centerline{\includegraphics{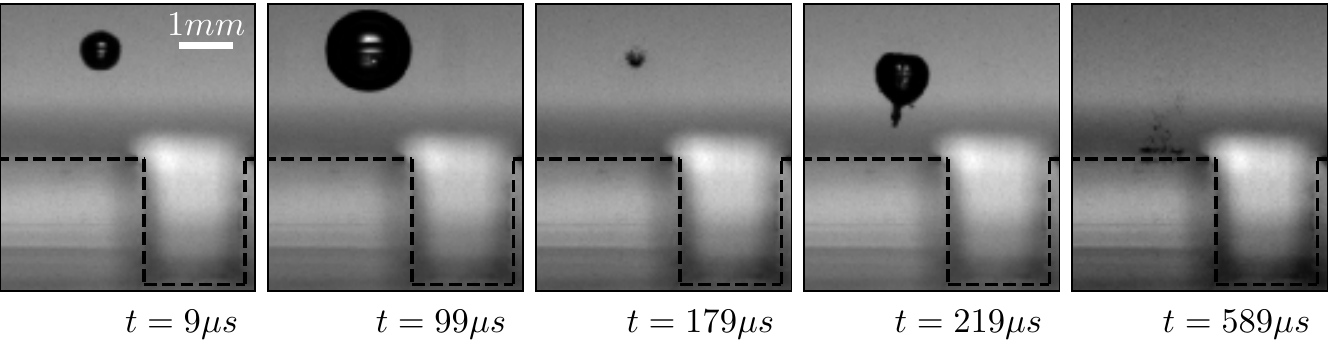}}
\caption{Snapshots from a high-speed recording of a bubble collapse near a slot with width $W = 2.2$ mm and $H = 2.7$ mm with the bubble positioned at a vertical distance $Y = 2.29$ mm and horizontal distance $X = -2.03$ mm. The jet angle is measured to be $\theta = -0.099$ radians ($-5.7$ degrees). The slot outline is indicated by the black dashed line.}
\label{fig:example_collapse}
\end{figure}

The physical mechanisms affecting the jet direction can be qualitatively understood. When the bubble collapses it draws in fluid from the surroundings. When the bubble collapses in infinite fluid, with no nearby boundaries, the fluid is drawn in completely symmetrically, leading to a spherically symmetrical collapse. If a solid boundary is present, the fluid cannot be drawn directly through the boundary so the boundary is `impeding' the flow \citep{Blake1983}. This impedance means that the fluid on the boundary side of the bubble is slower than on the open fluid side. For this reason the bottom of the bubble collapses slower than the top, creating an impulse towards the boundary known as the Kelvin impulse \citep{Blake1983}. Using this idea of relative impedance, some predictions can be made about how the bubble should behave near the slot. The slot contains fluid, so it should be easier for the bubble to draw fluid from the slot than from a solid boundary, effectively having a lower relative impedance. Thus, the fluid on the slot side of the bubble will move more quickly than on the solid boundary side. This means that the jet should be directed away from the slot. An example of a bubble collapse near a slot is shown in figure \ref{fig:example_collapse}; note how the jet is directed away from the slot. Based on symmetry and limiting behaviour, two more predictions can be made. First, based on symmetry, when the bubble is above the centre of the slot, the jet should be directed straight down ($\theta = 0$). Second, as the bubble moves infinitely far from the slot ($x \to \infty$) the boundary becomes a simple flat boundary and the jet must also be directed straight down. Based on these predictions, it is expected that there will be some maximum $\theta$ between $x=0$ and $x \to \infty$ with a negative minimum of an equal magnitude on the opposite side of the slot.

\section{Methods}
\subsection{Experimental Method}
\label{sec:Experimental Method}
\begin{figure}
\centerline{\includegraphics{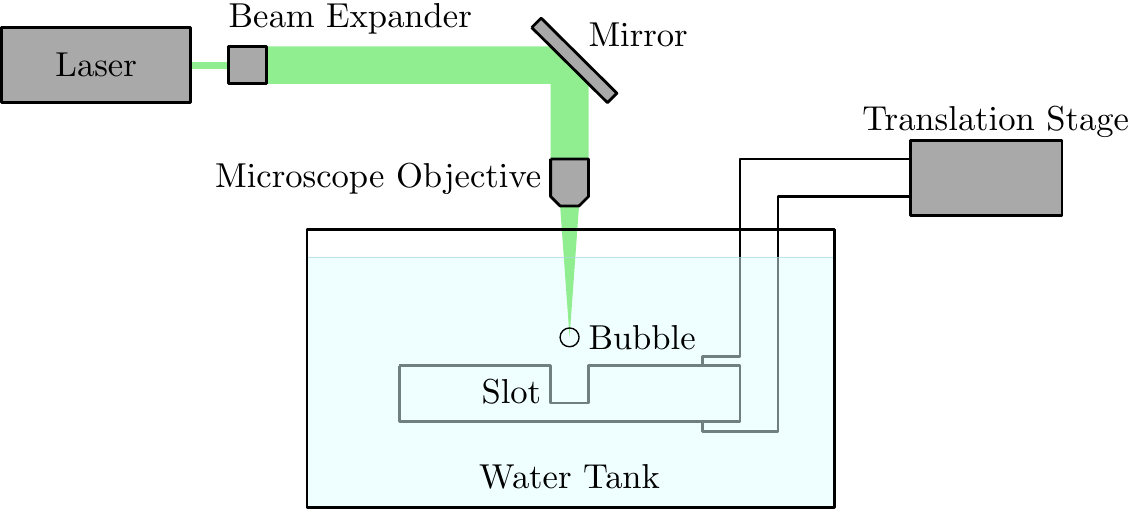}}
\caption{Diagram of the experiment configuration. In addition to the equipment shown, a high speed camera observes the bubble collapse, back-lit by a 100W LED panel.}
\label{fig:experiment_setup}
\end{figure}

Experiments were conducted using laser-induced cavitation \citep{Lauterborn1972, Noack1999}. Slot geometries for a range of slot widths and heights were created by layering laser cut acrylic, as shown in figure \ref{fig:slot_schematic}a. The experiment configuration is shown in figure \ref{fig:experiment_setup}. The slot geometry was connected to an arm supported by a translation stage. The translation stage was able to translate the geometry in three dimensions to an accuracy of 5 $\mu$m. The slot geometry was submerged approximately $25$ mm deep in a $180 \times 180 \times 100$ mm$^3$ acrylic tank of water. Prior to conducting experiments the water was degassed by subjecting it to a near vacuum in a vacuum chamber for approximately 30 minutes. A Q-switched Nd:YAG laser (`Bernoulli PIV' from Litron Lasers) was used to generate a 6 ns pulse at a wavelength of 532 nm. The laser output energy was set to 140 mJ per pulse, and then modulated down by adjusting the attenuator setting (typically between 30-35\%) to use the minimal level that could produce a bubble.  The laser pulse was passed through a beam expander as it left the laser and reflected downwards by a mirror. It then passed through a Nikon Plan Fluor 10X microscope objective (Numerical Aperture NA = 0.30) to focus the laser to the position where the bubble was created. The high numerical aperture, in combination with the expanded laser pulse, ensured that the laser could use minimal power and would not create any secondary bubbles along the laser path \citep{Sinibaldi2019}. A Photron FASTCAM SA-X2 high speed camera was used to record the bubble collapse. The camera recorded at 100 000 frames per second and used a 105 mm Nikon Micro-Nikkor lens. A 550 nm longpass filter was used to protect the camera from the laser. The camera triggered the laser to ensure accurate synchronising of the recording and the bubble collapse. For each bubble collapse, 100 frames were recorded, spanning 1 ms of time. A 100W LED panel was used to back-light the bubble collapse such that the bubble appears dark on a light background in the recorded images. Figure \ref{fig:example_collapse} shows snapshots from one such recording.

Recordings were post-processed in Python to measure key characteristics such as the bubble size variation over time and the direction of the resulting jet. Initially the background was removed from each frame of the recording, and a threshold filter was applied to isolate the bubble. The number of pixels provided a measure of bubble size, and the geometric centre of the pixels provided the position of the bubble centroid. To determine the jet direction, a vector was taken between the centroid positions of the first and second maxima of the bubble size (at 99 $\mu$s and 219 $\mu$s in figure \ref{fig:example_collapse}) which is generally a good approximation of the nominal jet direction at a position and is straightforward to measure \citep{PhysRevFluids.3.081601}. From this analysis, and position measurements from the translation stage, graphs were produced characterising how the jet direction varied with bubble position and geometry characteristics.

\subsection{Numerical Method}
\label{sec:Numerical Method}
The collapse of a gas or vapour bubble occurs on a timescale where viscous effects are not significant and the water can be considered incompressible for the majority of the collapse. This means that the collapse can be modelled by a velocity potential $\phi$. Various methods have been used to solve for this velocity potential. Simple potential flow models using mirror sinks have been used to predict the jet direction in limited geometries \citep{Kucera1990, PhysRevFluids.3.081601, Molefe2019}. More complex boundary integral methods and boundary element methods have been applied, typically to simulate how the surface of the bubble moves during collapse \citep{Kucera1990, Harris1996, Li2016, Brujan2019}. In this research a simplified boundary element method is used to predict the jet direction without resolving how the surface of the bubble moves over time. This is similar to the mirror sink model employed by \cite{PhysRevFluids.3.081601} and \cite{Molefe2019} to compute jet direction but with an infinite distribution of `mirror' sinks along the geometry boundary as in boundary element methods. This method combines the simplicity of the mirror sink model for finding the jet direction with the geometric versatility of boundary element methods. Using a simple model keeps the computational cost low, allowing for analysis of many different geometries across large parameter spaces. The method employed in this research is similar in form to that in \cite{Harris1996} but without solving for the bubble variation with time. Three key derivations are included here for completeness: the numerical method used (section \ref{sec:bem}), the derivation of the effect of a panel on itself (section \ref{sec:bem panel integral}), and the derivation of average bubble surface velocity for considering the effect of bubble size (section \ref{sec:average surface velocity}).

\subsubsection{Boundary Element Method}
\label{sec:bem}
The boundary is modelled as an infinite distribution of sinks on the boundary surface with varying strength. The infinite distribution of sinks is divided into panels such that the strength of the sinks within a panel can be approximated by a constant strength across the whole panel. The strength of each panel is determined by asserting no flow through the boundary surface at the centre of every panel and solving the resulting system of linear equations. The derivation for these equations is shown here.

A sink at position $\mathbf{x}_s$ induces a velocity $\nabla \phi = \mathbf{u}$ at any position $\mathbf{x} \neq \mathbf{x}_s$
\begin{equation}
\mathbf{u} = \frac{m (\mathbf{x} - \mathbf{x}_s)}{4 \pi | \mathbf{x} - \mathbf{x}_s | ^ 3},
\label{eq:velocity_from_sink}
\end{equation}
where $m$ is the volume flow rate of the sink, typically referred to as the sink strength.

The flow through a surface with normal vector $\mathbf{n}$ is the scalar product of the velocity at that point, $\mathbf{u}$, with the normal vector:
\begin{equation}
\frac{\partial \phi}{\partial n} = \frac{\partial \phi}{\partial x}n_x + \frac{\partial \phi}{\partial y}n_y + \frac{\partial \phi}{\partial z}n_z = \mathbf{u} \cdot \mathbf{n}.
\label{eq:normal_velocity}
\end{equation}

The velocity through the wall is a combination of the normal velocity from the bubble, $\partial\phi_b/\partial n$, and the normal velocity from the wall, $\partial \phi_w / \partial n$, which must sum to zero for the no-through-flow condition to be met:
\begin{equation}
\frac{\partial \phi}{\partial n} = \frac{\partial \phi_b}{\partial n} + \frac{\partial \phi_w}{\partial n} = 0.
\end{equation}

Thus, the normal component of the velocity from the wall can be expressed as the negative of the normal component of the velocity potential from the bubble.
\begin{equation}
\frac{\partial \phi_w}{\partial n} =  - \frac{\partial \phi_b}{\partial n}
\label{eq:wall_vel_negative_bubble_vel}
\end{equation}

To determine the normal velocity from the bubble at a position $\mathbf{x}$, the results from equations \ref{eq:velocity_from_sink} and \ref{eq:normal_velocity} can be combined to give
\begin{equation}
\frac{\partial \phi_b}{\partial n} = \mathbf{u}_b \cdot \mathbf{n} = \frac{m_b (\mathbf{x} - \mathbf{x}_b)}{4 \pi | \mathbf{x} - \mathbf{x}_b | ^ 3} \cdot \mathbf{n},
\label{eq:normal_bubble_velocity}
\end{equation}
where $\mathbf{x}_b$ is the position of the bubble.

Similarly, the velocity contribution from the wall can be expressed as the integral of the boundary sink strength density $\sigma$ over the wall surface $W$.
\begin{equation}
\frac{\partial \phi_w}{\partial n} = \iint\limits_W \sigma (\mathbf{x}_w) \frac{\mathbf{x} - \mathbf{x}_w}{4 \pi |\mathbf{x} - \mathbf{x}_w|^3} \cdot \mathbf{n} \mathrm{d}W,
\label{eq:integral}
\end{equation}
where $\mathbf{x}_w$ is a position on the wall surface.

The wall surface is expressed as a series of $N$ panels, each with a constant sink strength density, $\sigma_j$, centroid position, $\mathbf{x}_j$, and an area, $A_j$. Thus, equation \ref{eq:integral} is approximated by
\begin{equation}
\frac{\partial \phi_w}{\partial n} = \sum_{j=1}^N \sigma_j \frac{ A_j(\mathbf{x} - \mathbf{x}_j) \cdot \mathbf{n}}{4 \pi |\mathbf{x} - \mathbf{x}_j|^3}.
\label{eq:normal_velocity_summation}
\end{equation}

Substituting equations \ref{eq:normal_bubble_velocity} and \ref{eq:normal_velocity_summation} into \ref{eq:wall_vel_negative_bubble_vel} yields
\begin{equation}
\sum_{j=1}^N \sigma_j \frac{ A_j(\mathbf{x} - \mathbf{x}_j) \cdot \mathbf{n}}{4 \pi |\mathbf{x} - \mathbf{x}_j|^3} = - m_b\frac{ (\mathbf{x} - \mathbf{x}_b)\cdot \mathbf{n}}{4 \pi | \mathbf{x} - \mathbf{x}_b | ^ 3}.
\label{eq:full_linear_equation}
\end{equation}

This can be rewritten in terms of factors relating to the relative positions of each point. In this research, $R$ denotes these factors such that
\begin{equation}
R_{j} = \frac{ A_j(\mathbf{x} - \mathbf{x}_j) \cdot \mathbf{n}}{4 \pi |\mathbf{x} - \mathbf{x}_j|^3},
\label{eq:R_factor}
\end{equation}
\begin{equation}
R_{b} = \frac{(\mathbf{x} - \mathbf{x}_b) \cdot \mathbf{n}}{4 \pi |\mathbf{x} - \mathbf{x}_b|^3},
\label{eq:bubble_R_factor}
\end{equation}
and
\begin{equation}
\sum_{j=1}^N R_j \sigma_j = - m_b R_b.
\label{eq:simplified_linear_equation}
\end{equation}

In order to calculate the sink strength densities, $\sigma_j$, equation \ref{eq:simplified_linear_equation} must be solved at a number of points on the boundary equal to the number of panels. For a reasonable distribution of points, and convenience, these points are selected to be the panel centroids. Defining $R_{12}$ as the effect of the second panel at the centroid of the first panel, and so on, we write
\begin{equation}
\setlength{\arraycolsep}{0pt}
\renewcommand{\arraystretch}{1.3}
\left[
\begin{array}{cccc}
  R_{11}  &  R_{12}  &  \ldots  &  R_{1N} \\
  R_{21}  &  R_{22}  &  \ldots  &  R_{2N} \\
  \vdots  &  \vdots  &  \ddots  &  \vdots \\
  R_{N1}  &  R_{N2}  &  \ldots  &  R_{NN} \\
\end{array}  \right] \boldsymbol{\sigma}
= -m_b \left[
\begin{array}{c}
  R_{b1}  \\
  R_{b2}  \\
  \vdots  \\
  R_{bN}  \\
\end{array} \right],
\label{eq:matrix_linear_system}
\end{equation}
\begin{equation}
\mathsfbi{R}\boldsymbol{\sigma} = -m_b \mathbf{R}_b,
\label{eq:simplified_linear_system}
\end{equation}
where each term of the $\mathsfbi{R}$ matrix is
\begin{equation}
R_{ij} = \frac{ A_j(\mathbf{x}_i - \mathbf{x}_j) \cdot \mathbf{n}_i}{4 \pi |\mathbf{x}_i - \mathbf{x}_j|^3},
\label{eq:matrix_R_factor}
\end{equation}
which is undefined for any value $i = j$. This is resolved in section \ref{sec:bem panel integral}.

As with any system of linear equations, the system in equation \ref{eq:simplified_linear_system} could be solved in many ways. However, it is noted that the $\mathsfbi{R}$ matrix is fixed for any given geometry and is entirely independent of bubble position. Thus, if multiple bubble positions need to be evaluated for a single boundary geometry, a single inversion of the $\mathsfbi{R}$ matrix can be used to solve the system for all bubble positions using
\begin{equation}
\boldsymbol{\sigma} = -m_b \mathsfbi{R}^{-1}\mathbf{R}_b.
\label{eq:linear_system_solution}
\end{equation}
Although other methods would be faster for solving the system for a single position, this method is far more efficient when multiple bubble positions need to be solved.

This method generally performs well, as will be demonstrated in later sections, but is vulnerable to ill-conditioned systems in some cases. However, such systems can be identified using the condition number of the $\mathsfbi{R}$ matrix and avoided.

\subsubsection{Panel Integral}
\label{sec:bem panel integral}
The normal velocity induced by a panel at its own centroid is required in order to solve equation \ref{eq:matrix_linear_system}. For the majority of the panel this poses no issue; the velocity induced by a point that is not at the centroid is entirely tangential to the panel and so does not contribute to the normal velocity. However, the velocity induced by the centroid itself is undefined. Following the example of \cite{Brebbia2001}, the singularity can be resolved.

\begin{figure}
\centerline{\includegraphics{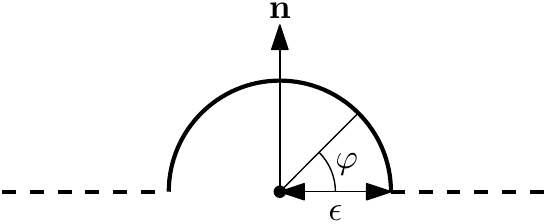}}
\caption{A cross-section of a panel (dashed line) deformed such that the centroid is a hemisphere (solid line), with radius $\epsilon$, centred on the panel centroid.}
\label{fig:panel_integral}
\end{figure}

The panel is deformed such that the centroid is expanded into a hemisphere with radius $\epsilon$, as shown in figure \ref{fig:panel_integral}. The velocity normal to the panel at the centroid becomes the integral of the normal velocity induced by the sink at the centroid position over the whole hemisphere.

For any point on the hemisphere, the velocity induced by the centroid sink is always normal to the surface. The normal velocity at any such point is thus
\begin{equation}
\mathbf{u}\cdot\mathbf{n} = \frac{\sigma}{4 \pi \epsilon^2}.
\end{equation}
For a circle on the hemisphere aligned parallel to the panel, with all points at an angle $\varphi$ from the horizontal, the sum of velocities is
\begin{equation}
u_{circle} = \frac{\sigma}{4 \pi \epsilon^2} 2 \pi \epsilon cos(\varphi) = \frac{\sigma}{2 \epsilon} cos(\varphi).
\end{equation}
This is integrated over the whole hemisphere to give
\begin{equation}
u_{total} = \int_0^{\pi / 2} \frac{\sigma}{2 \epsilon} cos(\varphi) \epsilon \mathrm{d}\varphi = \frac{\sigma}{2}  \int_0^{\pi / 2} cos(\varphi) \mathrm{d}\varphi = \frac{\sigma}{2},
\end{equation}
which is independent of the hemisphere radius, $\epsilon$. As $\epsilon \to 0$, the velocity induced by the panel at its own centroid is a constant $\sigma / 2$ and so the relative position factor is $R = 0.5$.

\subsubsection{Average Surface Velocity}
\label{sec:average surface velocity}
Previous research \citep{PhysRevFluids.3.081601, Molefe2019} has shown that the jet direction is accurately predicted by the velocity that the mirror sinks induce at the position of the bubble centroid.

The jet direction is aligned with the bubble translation velocity which is defined as the velocity of the centroid. We define the centroid as the average position of the bubble surface, so the velocity of the centroid is the average velocity of the bubble surface. In this section, it is shown that the induced velocity at the centroid of a spherical bubble is equal to the average velocity of the bubble surface, independent of bubble size. A more rigorous version of this derivation may be found in appendix \ref{sec:full average surface velocity derivation}.

The velocity induced by any individual sink on a point is given by
\begin{equation}
\mathbf{u} = -\frac{m\mathbf{r}}{4 \pi |\mathbf{r}|^3},
\end{equation}
where $\mathbf{r}$ is the vector from the sink to the point.

By integrating $\mathbf{u}$ over a sphere it can be shown that the total velocity on the surface of the sphere is
\begin{equation}
\mathbf{u}_{total} = -m \frac{\mathcal{R}^2\mathbf{d}}{|\mathbf{d}|^3},
\end{equation}
where $\mathcal{R}$ is the radius of the sphere and $\mathbf{d}$ is the vector from the sink to the bubble centroid.

The average velocity on the surface of the sphere is thus the total velocity divided by the surface area of the sphere,
\begin{equation}
\mathbf{u}_{avg} = \frac{\mathbf{u}_{total}}{4 \pi \mathcal{R}^2} = -\frac{m \mathcal{R}^2 \mathbf{d}}{4 \pi \mathcal{R}^2 |\mathbf{d}|^3} = -\frac{m \mathbf{d}}{4 \pi |\mathbf{d}|^3},
\end{equation}
which is equal to the velocity induced by the sink at the bubble centroid.

Due to the properties of potential flow, the velocity induced by every sink is simply a linear summation of the velocity induced by each individual sink. Thus, the average surface velocity of a sphere induced by a combination of sinks is equal to the velocity at the centre of the sphere. It is noted that the bubble sink itself does not contribute to the average surface velocity as all components cancel out.

\section{Results and Discussion}
\subsection{Numerical Results}
\label{sec:Numerical Results}
Using the boundary element method described in section \ref{sec:Numerical Method}, the jet direction can be modelled for given geometric parameters. Up to 20 000 panels were used to model the boundary, with a distribution such that there were more panels near the slot than towards the edges of the plate. The Python implementation of the boundary element method developed for this research runs on a standard desktop computer, where the primary limitation is the memory required to store the matrices.

\begin{figure}
  \centerline{\includegraphics{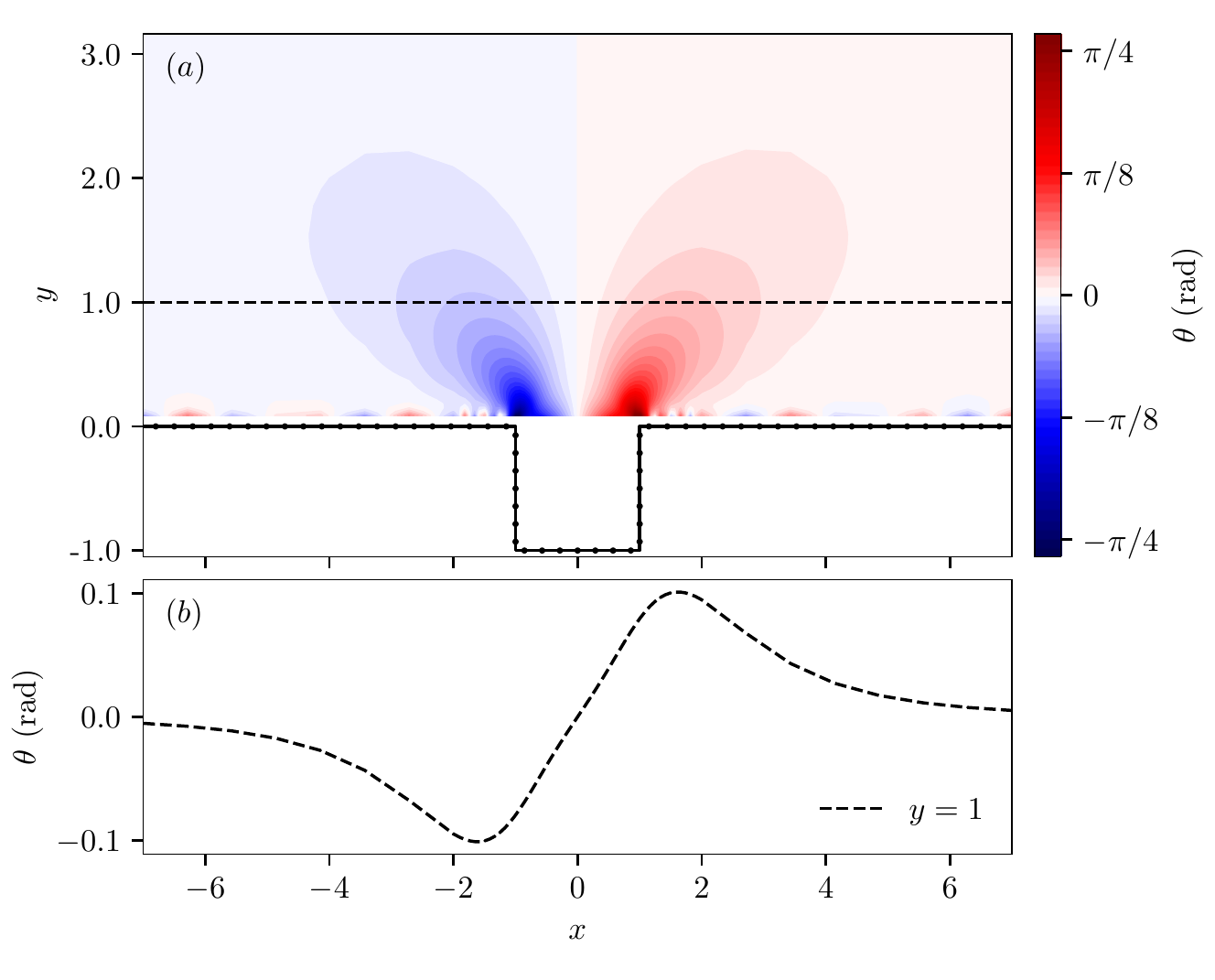}}
  \caption{(\textit{a}) A contour plot of $\theta$ as a function of $x$ and $y$. The slot, with $h = 1$, is represented by the black line, with points plotted at the panel centroids. Very close to the boundary the jet angle is more strongly affected by individual panel sinks rather than the boundary as a whole leading to the more rapid variations visible near the boundary. (\textit{b}) A plot of $\theta$ against $x$ for $y = 1$, corresponding to the black dashed line on the contour plot in (\textit{a}).}
\label{fig:jet_angle_contour}
\end{figure}

Figure \ref{fig:jet_angle_contour} shows a contour plot of the jet angle for a slot with a square cross-section ($h = 1$). As predicted qualitatively, the jet is directed towards the boundary but angled away from the slot. Three main regimes are revealed. Close to the boundary but far from the slot (low $y$, high $x$) the jet angle is dominated by the flat boundary and is not significantly affected by the slot. This leads to a very low jet angle. Near the centre of the slot the geometric symmetry dominates, again leading to low jet angles. The third regime is between these two, where the bubble is far enough from the centre that symmetry does not dominate, and positioned such that the slot has a significant effect relative to the flat boundary.

In order to more easily quantify this shape, and for simpler comparison to experimental results, slices of the $\theta$ contour are taken with constant $y$ and a range of $x$. One such slice is shown in figure \ref{fig:jet_angle_contour}b and is the form in which data will be presented hereafter. As in the contour plot, this plot shows that the jet tends to be directed straight at the boundary when the bubble is far from the slot. As the bubble approaches the slot the jet is angled away from the slot centre down to a negative peak jet angle. After the peak the jet angle tends back towards zero, crossing zero at the centre point. The jet angle then increases to an equal and opposite peak on the other side of the slot due to the symmetry of the geometry.

\begin{figure}
  \centerline{\includegraphics{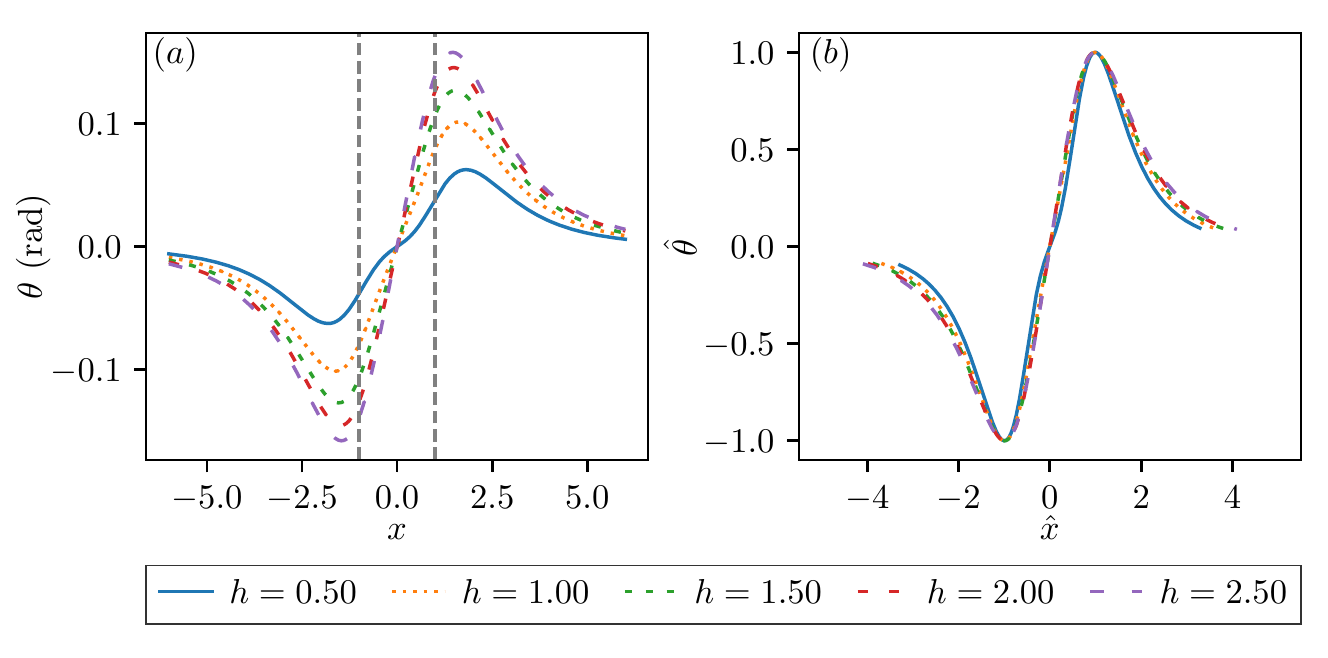}}
  \caption{(\textit{a}) $\theta$ against $x$ for a range of slot geometries characterised by $h$ and a fixed value of $y = 1$. (\textit{b}) Normalised $\theta$, $\hat{\theta}$, against normalised $x$, $\hat{x}$, for the same values of $h$. The curves collapse well onto one curve when normalised. The $h = 1$ curves here are the same as the $y = 1$ curves in figure \ref{fig:y_collapse}.}
\label{fig:h_collapse}
\end{figure}

\begin{figure}
  \centerline{\includegraphics{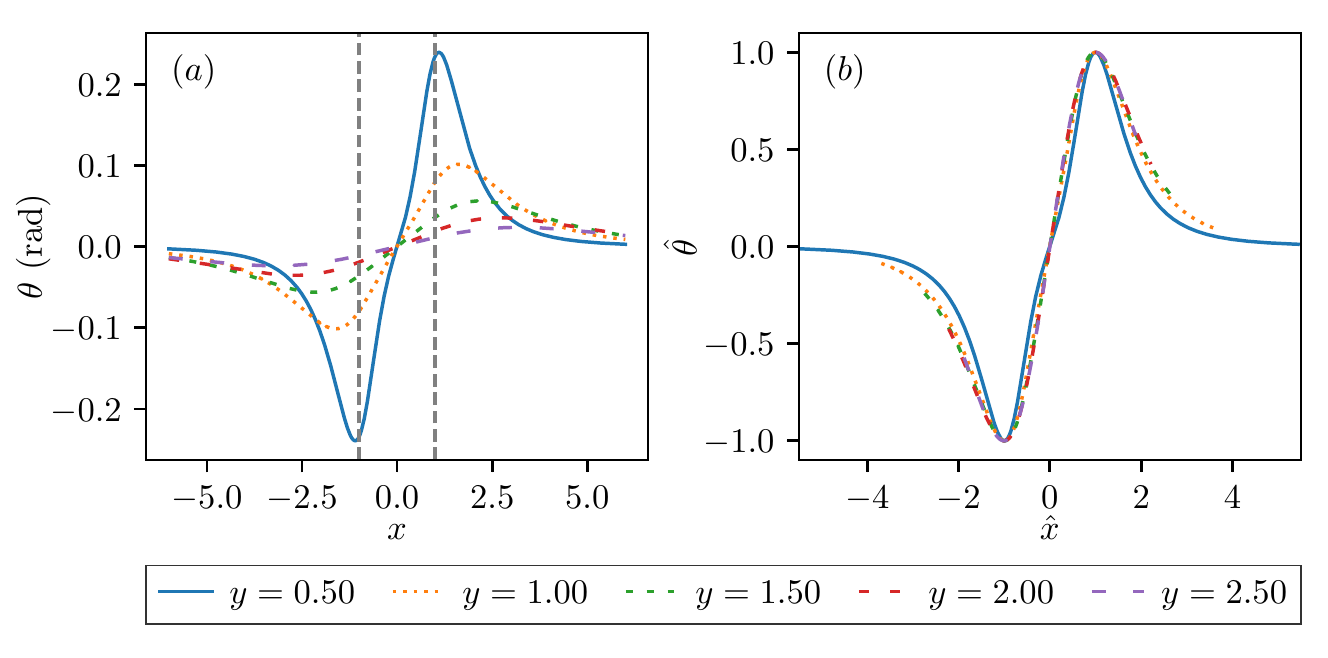}}
  \caption{(\textit{a}) $\theta$ against $x$ for a range of slot geometries characterised by $y$ and a fixed value of $h = 1$. (\textit{b}) Normalised $\theta$, $\hat{\theta}$, against normalised $x$, $\hat{x}$, for the same values of $y$. The curves collapse well onto one curve when normalised. The $y = 1$ curves here are the same as the $h = 1$ curves in figure \ref{fig:h_collapse}.}
\label{fig:y_collapse}
\end{figure}

Using the normalised geometric parameters, two sets of numerical predictions have been plotted in figures \ref{fig:h_collapse} and \ref{fig:y_collapse} to show how the jet angle curve varies with $h$ and $y$.

Figure \ref{fig:h_collapse}a shows that the peak jet angle increases as the normalised slot height, $h$, increases. This is because the slot contains more fluid and so its relative impedance is decreased. The position of the peak moves closer to the slot as the height is increased. These curves have a similar shape, which can be characterised by two parameters defining the peak. For a given $y$ and $h$, the maximum value of $\theta$ is $\theta^\star$ and occurs at $x = x^\star$. Thus, $\theta$ can be normalised with $\theta^\star$, and $x$ can be normalised with $x^\star$.
\begin{equation}
\hat{\theta} = \frac{\theta}{\theta^\star}, \hat{x} = \frac{x}{x^\star}
\label{eq:normalised_theta_and_p}
\end{equation}
When these curves are normalised, as shown in figure \ref{fig:h_collapse}b, they collapse down to being very close to the same curve, with the exception of the data with the lowest height ($h = 0.5$).

Figure \ref{fig:y_collapse}a shows that the peak jet angle, $\theta^\star$, increases as the normalised vertical distance away, $y$, decreases. This is because the bubble is closer to the boundary and so is more strongly affected. The position of the peak, $x^\star$, moves further from the slot as the vertical distance away is increased. When these curves are normalised, as shown in figure \ref{fig:y_collapse}b, they also collapse down to being very close to the same curve, with the exception of the data closest to the boundary ($y = 0.5$).

It is noted that the $h = 1$ curves in figure \ref{fig:h_collapse} are the same as the $y = 1$ curves in figure \ref{fig:y_collapse} showing that the curves from both figures collapse to the same curve.

As mentioned, the curves collapse except for bubbles very close to the boundary, or with slots that have a low height. For these cases, the effect of the slot can be treated as two opposing steps with a relatively large separation. A single step would cause a single jet angle peak. When the effects of two peaks with opposite signs are combined, the gradient in the middle depends on the separation of the peaks. If the peaks are very close,  then the peaks blend together resulting in the collapsed shapes seen in figures \ref{fig:h_collapse}b and \ref{fig:y_collapse}b. If, however, the peaks are far apart then a pronounced kink in the curve appears at the middle of the slot, as shown by the $h = 0.5$ curve in figure \ref{fig:h_collapse} and the $y = 0.5$ curve in figure \ref{fig:y_collapse}. When bubbles are close to the boundary the effects of each side of the slot are relatively more separated, as is also the case for low height slots. Most slots have $h \geq 1$ and $y \geq 1$ where the peaks merge giving the collapsed curve shape; only these curves will be considered in further analysis.

\begin{figure}
  \centerline{\includegraphics{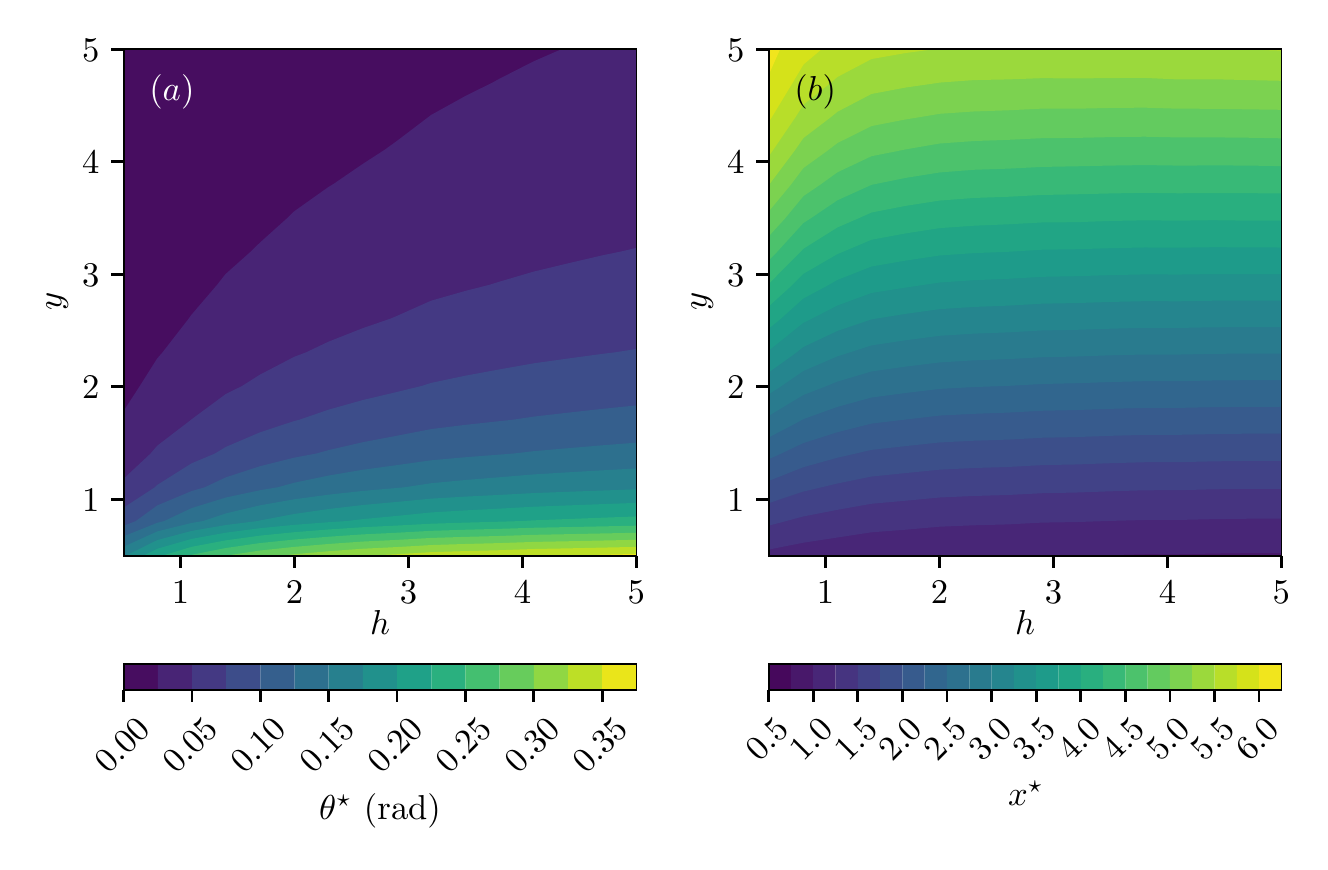}}
  \caption{Contour plots of peak jet angle, $\theta^\star$, and peak jet angle position, $x^\star$, as functions of non-dimensional vertical distance, $y$, and non-dimensional slot height, $h$.}
  \label{fig:peak_variation_contours}
\end{figure}

As the curves in figures \ref{fig:h_collapse} and \ref{fig:y_collapse} collapse well onto the same curve, $\theta^\star$ and $x^\star$ can be used to characterise the variation of the jet angle with the parameters $h$ and $y$. Two contour plots in figure \ref{fig:peak_variation_contours} show $\theta^\star$ and $x^\star$ plotted as functions of $h$ and $y$.

Figure \ref{fig:peak_variation_contours}a shows that the maximum jet angle, $\theta^\star$, depends most strongly on the dimensionless vertical bubble position, $y$. As $y$ increases, with the bubble moving far from the slot, the maximum jet angle tends towards zero. As $y$ tends towards zero, the maximum jet angle increases very rapidly, but is bounded by the solution for a convex right-angle corner as would be the case for a bubble at $y = 0$ as $x \to 1^-$.

Figure \ref{fig:peak_variation_contours}a shows a weaker dependence of maximum jet angle on the dimensionless slot height, $h$, compared to the dependence on $y$. As $h$ increases, the maximum jet angle tends towards a limit, as can be observed in the contour, showing only a dependence on $y$ for large values of $h$. The figure also shows that the maximum jet angle tends to zero as $h$ tends to zero which is expected as the slot becomes a flat plate.

Figure \ref{fig:peak_variation_contours}b shows the dependence of the maximum jet angle, $x^\star$, on both $y$ and $h$. $x^\star$ increases approximately linearly with $y$, as might be predicted from figure \ref{fig:jet_angle_contour}a. For low values of $h$ the gradient of the linear relationship changes, but as $h$ increases the gradient tends towards $1$ (this can be seen in figure \ref{fig:peak_variation_comparison}b where the gradient of the $x^\star$-$y$ line is approximately $1$). It is noted that, for very low jet angles, the position of the peak becomes more sensitive to numerical errors and imperfect boundary conditions, such as the edges of the plate. Very low jet angles are found at low $h$ and high $y$ values, in the top left corners of the contour plots in figure \ref{fig:peak_variation_contours}.

Both $\theta^\star$ and $x^\star$ exhibit limiting behaviour as $h$ increases. This suggests that at some point increasing the height of the slot will have a negligible effect on the jet angle. Even for lower values of $h$, the variation with $h$ is typically much less significant than with $y$.

\subsection{Experimental Results}
\label{sec:Experimental Results}
Experiments were conducted by performing a horizontal sweep over a selected slot at fixed vertical distances $Y$. The horizontal positions tested were selected to focus most of the data around both $\theta$ peaks and to observe the behaviour at a large horizontal distance on at least one side of the slot. Each position was tested multiple times; for most experiments there were five repeats.

In order to understand the stochastic variation of $\theta$ that occurs at each position, two series of experiments were conducted with 50 repeats at each position. From this data it was observed that the standard deviation of jet angle is reasonably consistent for all positions and conforms well to the normal distribution. This standard deviation was therefore applied to the remainder of the data to provide statistical error bars that show a 99\% confidence interval of the mean at each position based on the number of repeats at those positions.

A second order polynomial curve fit was applied around each of the two peaks of the $\theta$-$x$ curve. As the geometry is symmetrical the two peaks should be equal and opposite; the peak on the negative $x$ side should have a negative $\theta$ value of the same magnitude as on the positive $x$ side. Thus, if the polynomial curve fits on each side are slightly offset from being symmetrical, all of the data can be shifted by the offset in both $\theta$ and $x$ to achieve symmetry. An offset in the $\theta$ axis could, for example, be caused by a tilted camera frame. In addition, the curve fit from one side, mirrored in both axes, should fit the data from the opposite side. If this is not the case then it can be concluded that the data is of a low quality and thus neglect it from further analysis. Of the data collected in this research, horizontal sweeps at three $y$ values for the W2H6 geometry were neglected on this basis and thus not presented here.

\begin{figure}
\centerline{\includegraphics{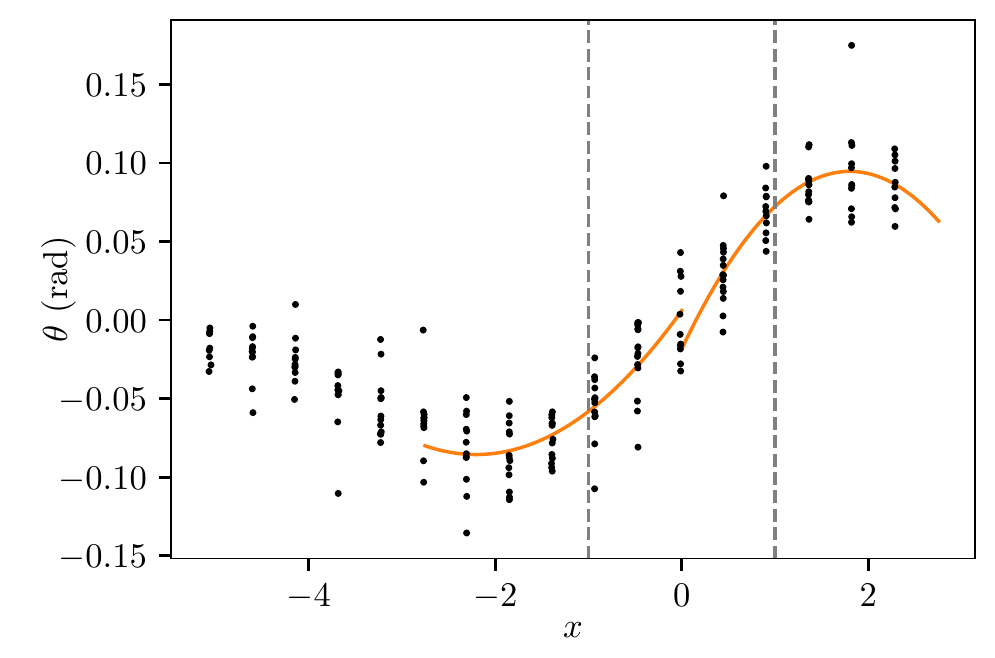}}
\caption{An example curve fit on data using slot `W2H3a' from table \ref{tab:geometries} ($W = 2.2$ mm, $H = 2.7$ mm). The bubble was positioned at a vertical distance $Y = 2.81$ mm. The points are the experimental data and the lines are the two curve fits.}
\label{fig:curve_fit}
\end{figure}

Data was gathered for both peaks so that this symmetry analysis could be conducted. However, only one side of the slot was tested to a greater horizontal distance as the behaviour in this region is already well understood. An example sweep with the curve fit plotted is shown in figure \ref{fig:curve_fit}. In this example the peaks are slightly offset from being symmetrical, so the data would be shifted before conducting further analysis.

\begin{figure}
  \centerline{\includegraphics{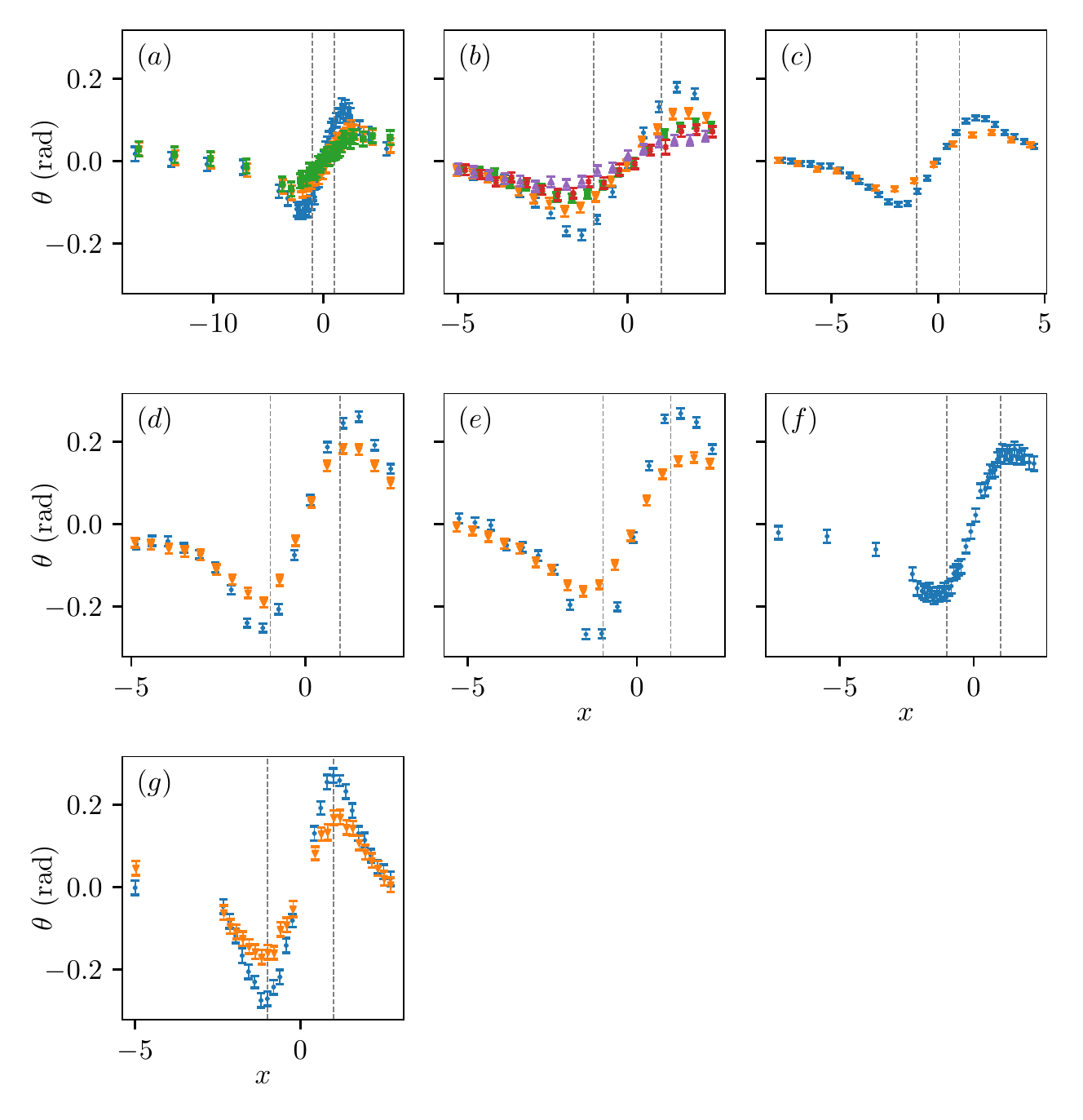}}
  \caption{$\theta$ plotted against $x$ for all experimental data from table \ref{tab:geometries}. The slot boundaries are indicated by the dashed vertical lines ($x = -1$, $x = 1$). Data is coloured by $y$ value; peak $\theta$ decreases as $y$ increases. ($a$) W1H3: blue $y = 1.58$, orange $y = 2.37$, green $y = 3.16$. ($b$) W2H3a: blue $y = 0.80$, orange $y = 1.04$, green $y = 1.28$, red $y = 1.51$, purple $y = 1.75$. ($c$) W2H3b: blue $y = 1.21$, orange $y = 1.67$. ($d$) W2H6: blue $y = 0.69$, orange $y = 0.90$. ($e$) W2H9: blue $y = 0.78$, orange $y = 1.24$. ($f$) W2H12: blue $y = 1.20$. ($g$) W4H12: blue $y = 0.58$, orange $y = 0.82$.}
  \label{fig:experiment_data_panels}
\end{figure}

The geometries tested are shown in table \ref{tab:geometries} and the jet angle data is summarised in figure \ref{fig:experiment_data_panels}. The experimental results follow the same qualitative trends as the numerical predictions. There is a negative jet angle peak on the left side of the slot, and a positive jet angle peak on the right. The magnitude of jet angle peak increases as $y$ decreases and $h$ increases. The position of the jet angle peak increases as $y$ increases, but it is more difficult to discern how the position varies with $h$.

\begin{table}
  \begin{center}
\def~{\hphantom{0}}
  \begin{tabular}{lccccc}
      Label &  $W$ (mm)    &   $H$ (mm)   &   $Y$ values measured (mm)        &   $h$    &   $y$                          \\[3pt]
      W1H3  &  1.23        &   ~2.74      &   1.94, 2.91, 3.89                &   2.23   &   1.58, 2.37, 3.16             \\
      W2H3a &  2.20        &   ~2.70      &   1.77, 2.29, 2.81, 3.32, 3.84    &   1.23   &   0.80, 1.04, 1.28, 1.51, 1.75 \\
      W2H3b &  2.20        &   ~2.90      &   2.66, 3.68                      &   1.32   &   1.21, 1.67                   \\
      W2H6  &  2.20        &   ~5.40      &   1.52, 1.99                      &   2.45   &   0.69, 0.90                   \\
      W2H9  &  2.14        &   ~8.21      &   1.66, 2.66                      &   3.84   &   0.78, 1.24                   \\
      W2H12 &  2.20        &   11.50      &   2.63                            &   5.23   &   1.20                         \\
      W4H12 &  4.20        &   11.47      &   2.43, 3.43                      &   2.73   &   0.58, 0.82                   \\
  \end{tabular}
  \caption{Measurements of geometries used. Labels refer to the nominal width and height.}
  \label{tab:geometries}
  \end{center}
\end{table}

\subsection{Comparison}
\label{sec:Comparison}
We will now proceed to directly compare our experimental data to the numerical results.

\begin{figure}
  \centerline{\includegraphics{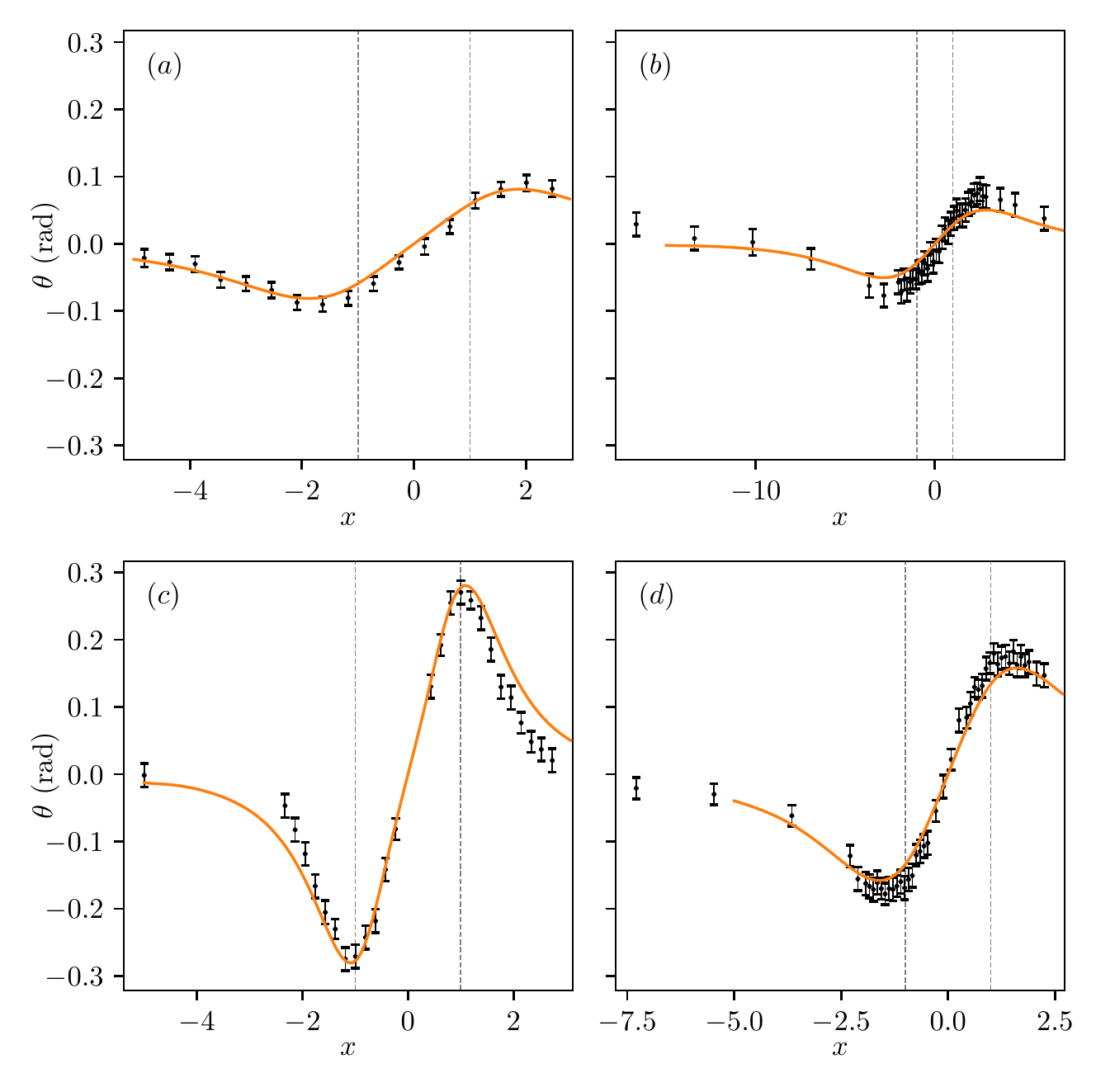}}
  \caption{Experimental data for a series of slots from table \ref{tab:geometries}. This data is compared to numerical predictions for the same configurations. Experimental data is plotted as points with error bars, numerical model predictions are plotted as solid lines. Geometries are ordered by $h$ value. ($a$) W2H3a, $h = 1.23$, $y = 1.28$. ($b$) W1H3, $h = 2.23$, $y = 2.37$. ($c$) W4H12, $h = 2.73$, $y = 0.58$. ($d$) W2H12, $h = 5.23$, $y = 1.20$.}
\label{fig:direct_comparison}
\end{figure}

Figure \ref{fig:direct_comparison} shows a direct comparison between four experimental data curves and boundary element method predictions for the same geometric parameters. These plots show a good agreement between predicted curves and experimental results. The most significant difference is for the W4H12 geometry where the experimental data has a steeper gradient on the outer sides of the peaks, but the magnitude and position of the peak is well predicted. It is noted that the W4H12 data in general has a steeper peak curve than other experimental data when compared to numerical results. There is a tendency for the numerical model to under-predict the magnitude of the jet angle peak, although the peak position is generally predicted well. On average, across all 17 horizontal sweeps presented here, the numerical model under-predicted $\theta^\star$ by $13$ \% and $x^\star$ by $2.6$ \%.

\begin{figure}
  \centerline{\includegraphics{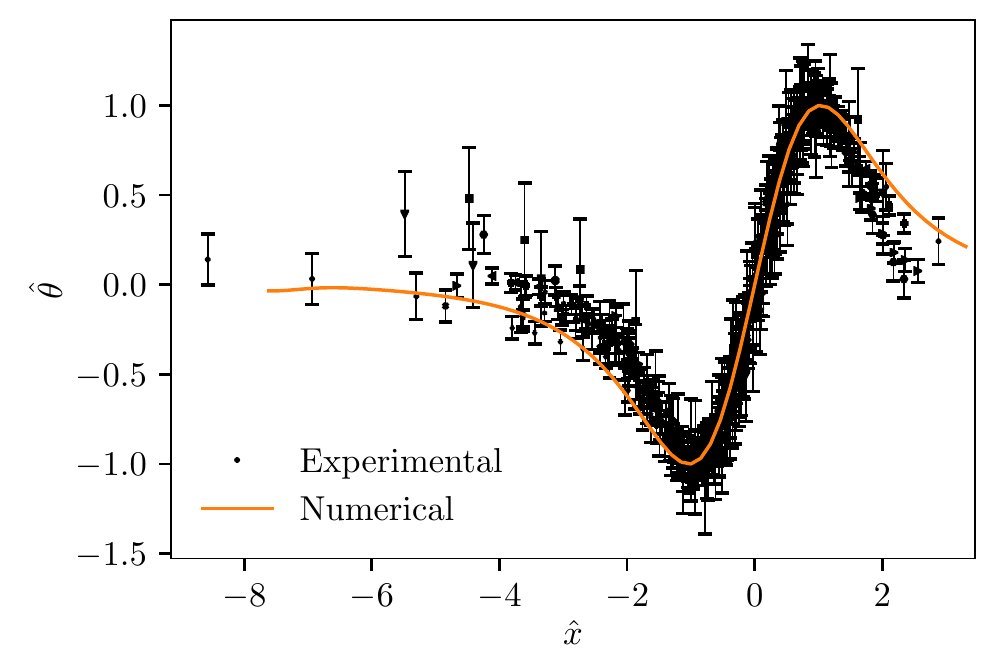}}
  \caption{Experimental results from all geometries, each with a range of $y$, showing normalised jet angle, $\hat{\theta}$, against normalised horizontal position, $\hat{x}$. This is compared to the collapsed numerical curve that matches all numerical curves for $y \geq 1$ and $h \geq 1$. The experimental results collapse down onto one curve and match the numerical prediction.}
\label{fig:all_data_collapse_comparison}
\end{figure}

All experimental data collapses onto a single curve when the experimental data is normalised with the peak values $x^\star$ and $\theta^\star$, which were determined using the previously described curve fitting method. Figure \ref{fig:all_data_collapse_comparison} shows all of the data normalised and compared to the normalised prediction curve that matches all predictions with $h \geq 1$ and $y \geq 1$. The experimental data collapses very well, validating the collapse observed from the numerical predictions. Although the numerical prediction curve has a slightly higher normalised jet angle on the outer sides of the peaks, the collapsed experimental data curve matches the numerical curve remarkably well. The variations in the normalised curve observed numerically at very low $h$ and at very low $y$ would likely not be visible in the experimental data due to the magnitude of the error compared to the magnitude of the variations.

\begin{figure}
  \centerline{\includegraphics[scale=0.75]{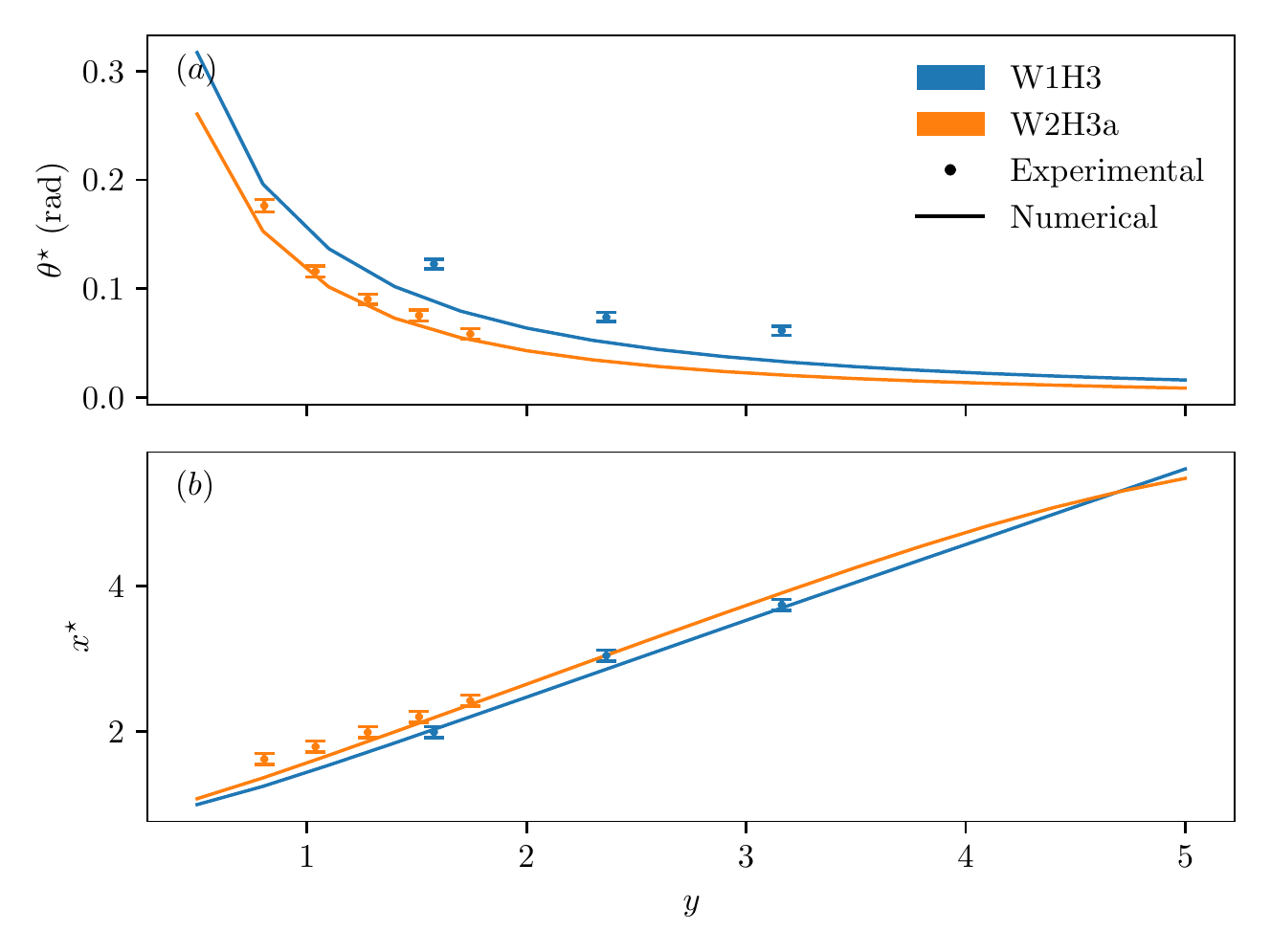}}
  \caption{Experimental data points from two slots compared with boundary element method predictions for the same configurations. (\textit{a}) Peak jet angle, $\theta^\star$, as a function of $y$. (\textit{b}) Position of peak jet angle, $x^\star$, as a function of $y$. Slot W1H3 has $h = 2.23$. Slot W2H3a has $h = 1.23$. Experimental data is plotted as dots, numerical model predictions are solid lines.}
\label{fig:peak_variation_comparison}
\end{figure}

A comparison between numerically predicted $\theta^\star$ and $x^\star$ trends and experimental results is shown in figure \ref{fig:peak_variation_comparison}. Here $\theta^\star$ and $x^\star$ have been calculated using the curve fitting method described in section \ref{sec:Experimental Results}. The error bars in this figure are based on the error distribution from large amounts of data, synthesised using the numerical model, with similar properties to experimental data. It is also noted from the synthesised data that the curve fit has a tendency to over-estimate the peak position. The results are compared for the same $h$ values as the experimental data and the same range of $y$ values as figure \ref{fig:peak_variation_contours}. These results generally show a good agreement between the numerical and experimental results. The most significant difference is that the numerical prediction underestimates the $\theta^\star$ values for W1H3. This is to be expected as a similar discrepancy is observed in figure \ref{fig:direct_comparison}.

Overall, the model tends to under-predict the jet angle, but performs well on the curve shape and trends. The position of the maximum jet angle, $x^\star$, is especially well predicted.

\section{Conclusion}
We have investigated the collapse of bubbles near a slot geometry using experiments and a simple numerical model. Our main observation is the variation of the jet angle with the horizontal position of the bubble. At the center of the slot the jet is directed straight downwards due to symmetry and far from the slot the surface acts as a flat boundary so the jet is also directed downwards. Between these two limits there is a peak jet angle deflection, angled away from the slot. The peak jet angle and position of the peak jet angle both tend to a limiting value as the slot height increases. This shows that, for slots of sufficient height, the jet angle depends only on the bubble position. As the vertical position of the bubble increases, the peak jet angle decreases and the horizontal position of the peak increases. For $h >> 1$, the position of the peak jet angle is directly proportional to the vertical distance: $x^\star \propto y$. When the jet angle and horizontal position are normalised by their respective peak values, we find that all jet angle curves collapse.

The numerical model has a tendency to under-predict the jet angle compared to experimental data gathered in this research, although it is often within reasonable error. The numerical model predictions very closely follow the shape and overall trends of the data, and provide a good prediction of the position of the peak jet angle, $x^\star$. In addition, the collapsed curve predicted by the numerical model matches the collapsed curve found from the experimental data. These comparisons serve to validate the numerical model presented in this research and provide a good basis from which to continue the study of complex geometries using this model. The velocity profile of the fluid in and around the slot can also be predicted by this model, although further investigation would be required to validate the predicted profiles.

In the context of cleaning with bubbles, particularly with ultrasonic cavitation, this research suggests that slots in surfaces to be cleaned would likely experience fewer jet impacts because bubbles would be drawn away from the slots rather than into them, and thus be cleaned less rigorously than the rest of the surface. Where cavitation damage is a problem this property could also be used to protect sensitive components by recessing them within slots, although this technique would likely have other implications depending on the flow conditions and requirements of the components.

\section*{Acknowledgements}
Many thanks to Lebo Molefe for invaluable discussions and comments. We acknowledge financial support from the EPSRC under Grant No. EP/P012981/1. DFR acknowledges the funding from the European Research Council (ERC) under the European Union’s Horizon 2020 research and innovation programme (Grant agreement No. 851630).

\section*{Declaration of Interests}
The authors report no conflict of interest.

\appendix
\section{Full Average Surface Velocity Derivation}
\label{sec:full average surface velocity derivation}
The translation velocity of the bubble is defined as the average velocity of the bubble surface and can be used to calculate the jet direction as the jet direction follows the translation of the bubble. However, many evaluations of the velocity field would be required to determine the average velocity at the bubble surface numerically, so a different method is preferred. From the method employed in \cite{PhysRevFluids.3.081601} and \cite{Molefe2019}, and from numerical testing, it is found that the velocity induced by mirror sinks at the bubble centre accurately describes the jet direction, independent of bubble radius. It is of interest to more fully understand this relationship and how it relates to the average surface velocity.

The potential flow model is a linear summation of multiple sinks, including the bubble and the wall or mirror sinks. Thus, the average velocity on the bubble surface for the whole system can be described as the summation of the average velocity on the bubble surface for each individual sink. It is assumed that the bubble surface is a perfect sphere, centred on the bubble position, with a radius $\mathcal{R}$. The average velocity on the bubble surface induced by its own sink is zero as each point has an equal and opposite point that sum to zero so the bubble sink can be neglected from further analysis.

\begin{figure}
\centerline{\includegraphics{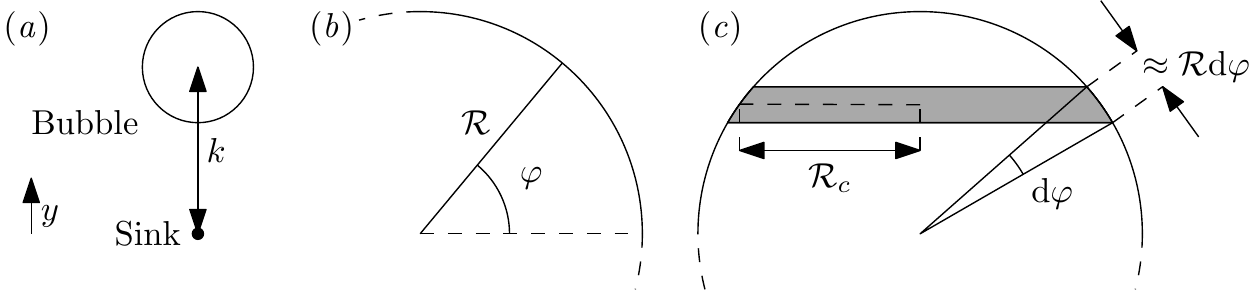}}
\caption{(\textit{a}) The arrangement of the bubble, sink, and the direction of the $y$ axis. (\textit{b}) The definition of $\varphi$ and $\mathcal{R}$. (\textit{c}) A section of the sphere surface used for integration.}
\label{fig:surface_integral_section}
\end{figure}

The velocity induced by a sink on an arbitrary point is directed at the sink and has a calculable magnitude. In the present analysis each sink is evaluated in isolation, so it can be assumed that the sink is positioned at the origin and the bubble is positioned at a distance $k$ along an axis from the sink to the bubble centre, as shown in figure \ref{fig:surface_integral_section}a. This reduces the problem to an axisymmetric configuration where the off-axis components of the velocity will cancel out. Thus, only the velocity in the axial direction need be evaluated. For this analysis the axis is denoted by $y$ and the axes normal to it are denoted by $x$ and $z$ in the usual manner. It is assumed that the bubble radius, $\mathcal{R}$, is less than the distance from the sink, $|\mathbf{r}|$, such that the bubble is not in contact with the sink. The velocity along the $y$ axis is
\begin{equation}
u_y = -\frac{m}{4 \pi} \frac{y}{|\mathbf{r}|^3},
\label{eq:vel_y_component}
\end{equation}
where $m$ is the sink strength, $y$ is the $y$ position of a point to be evaluated, and $\mathbf{r}$ is the position vector of the point to be evaluated. 

For this derivation $\varphi$ is defined as the angle anticlockwise from the horizontal towards the $y$ axis, as shown in figure \ref{fig:surface_integral_section}b. At $\varphi=-\pi/2$ the point is at the bottom of the sphere, and at $\varphi=\pi/2$ it is at the top of the sphere. Thus, for a point on the surface of the bubble, the $y$ position is defined as 
\begin{equation}
y = k + \mathcal{R} sin(\varphi).
\label{eq:y_position}
\end{equation}

The magnitude of the $\mathbf{r}$ vector is
\begin{eqnarray}
|\mathbf{r}| & = & \sqrt{y^2 + \mathcal{R}^2 cos^2(\varphi)} \nonumber \\
             & = & \sqrt{k^2 + 2k\mathcal{R}sin(\varphi) + \mathcal{R}^2sin^2(\varphi) + \mathcal{R}^2cos^2(\varphi)} \nonumber \\
             & = & \sqrt{k^2 + \mathcal{R}^2 + 2k\mathcal{R}sin(\varphi)}.
\label{eq:r_magnitude}
\end{eqnarray}
Note that $\mathcal{R}^2cos^2(\varphi)$ includes both the $x$ and $z$ components of the $r$ vector.

Thus, equation \ref{eq:vel_y_component} can be expanded as
\begin{equation}
u_y = - \frac{m}{4 \pi} \frac{k + \mathcal{R}sin(\varphi)}{(k^2 + \mathcal{R}^2 + 2k\mathcal{R}sin(\varphi))^{3/2}}.
\end{equation}

The sphere surface can be broken into a series of circles, aligned parallel to the $xz$ plane. The sum of the velocities around any such circle is described by
\begin{equation}
u_{yc} = 2 \pi \mathcal{R}_c u_y = 2 \pi \mathcal{R} cos(\varphi) u_y,
\label{eq:velocity_on_circle}
\end{equation}
where $\mathcal{R}_c = \mathcal{R}cos(\varphi)$ is the radius of the circle.

By considering a section of the sphere surface, as shown in figure \ref{fig:surface_integral_section}c, with a radius $\mathcal{R}$ and height $\mathcal{R}d\varphi$, equation \ref{eq:velocity_on_circle} can be integrated over the whole sphere to find the total velocity $u_{ytotal}$.
\begin{eqnarray}
u_{ytotal} & = & \int_{-\pi / 2}^{\pi / 2} u_{yc} \mathcal{R} \mathrm{d}\varphi \\
           & = & - \frac{m}{2} \mathcal{R}^2 \int_{-\pi / 2}^{\pi / 2} \frac{cos(\varphi) (k + \mathcal{R}sin(\varphi))}{(k^2 + \mathcal{R}^2 + 2k\mathcal{R}sin(\varphi))^{3/2}} \mathrm{d}\varphi \\
           & = & - \frac{m}{2} \mathcal{R}^2 \frac{\mathcal{R} + ksin(\varphi)}{k^2 \sqrt{k^2 + \mathcal{R}^2 + 2k\mathcal{R}sin(\varphi)}} \Big|_{-\pi / 2}^{\pi / 2} \\
           & = & - \frac{m}{2} \mathcal{R} ^2 \frac{2}{k^2} = - m \frac{\mathcal{R}^2}{k^2}
\label{eq:total_surface_velocity}
\end{eqnarray}

The average velocity is thus the total velocity $u_{ytotal}$ divided by the total area $4 \pi \mathcal{R}^2$. Substituting in the expression from equation \ref{eq:total_surface_velocity} yields 
\begin{equation}
u_{yavg} = \frac{u_{ytotal}}{4 \pi \mathcal{R}^2} = -\frac{m \mathcal{R}^2}{4 \pi \mathcal{R}^2 k^2} = - \frac{m}{4 \pi k^2},
\label{eq:average_surface_velocity}
\end{equation}
which is notably independent of the bubble radius $\mathcal{R}$. Equation \ref{eq:average_surface_velocity} is the same as the $y$ velocity evaluated at the centre of the bubble, found by substituting $y = k$ and $|\mathbf{r}| = k$ into equation \ref{eq:vel_y_component}.

Thus, the velocity induced at the centre of the bubble is equal to the average surface velocity. This can be extended to the whole system where the average surface velocity is equal to the velocity induced at the centre of the bubble by all sinks.
\begin{equation}
\mathbf{u} = \sum_{k=1}^N \sigma_k \frac{ A_k(\mathbf{x}_b - \mathbf{x}_k)}{4 \pi |\mathbf{x}_b - \mathbf{x}_k|^3}
\label{eq:bubble_position_velocity}
\end{equation}
Only the direction of $\mathbf{u}$ is considered, as the magnitude is effectively arbitrary in this research.

\bibliographystyle{jfm}
\bibliography{slot-geometries}
\end{document}